\documentclass[twocolumn,showpacs,preprintnumbers,amsmath,amssymb, prl]{revtex4}

\usepackage{graphicx}% Include figure files
\usepackage{dcolumn}% Align table columns on decimal point
\usepackage{bm}% bold math

\begin{document}

\preprint{}

\title{Suppression of electron spin relaxation in Mn-doped GaAs}

\author{G.~V.~Astakhov$^{1}$}
\altaffiliation{Also at A.~F.~Ioffe Physico-Technical Institute,
RAS, 194021 St. Petersburg, Russia} \email[\\ E-mail:
]{astakhov@physik.uni-wuerzburg.de}
\author{R.~I.~Dzhioev,$^{2}$}
\author{K.~V.~Kavokin$^{2}$}
\author{V.~L.~Korenev$^{2}$}
\author{M.~V.~Lazarev$^{2}$}
\author{M.~N.~Tkachuk $^{2}$}
\author{Yu.~G.~Kusrayev $^{2}$}
\author{T.~Kiessling$^{1}$}
\author{W.~Ossau$^{1}$}
\author{L.~W.~Molenkamp$^{1}$}

\affiliation{$^{1}$Physikalisches Institut (EP3), Universit\"{a}t
W\"{u}rzburg, 97074 W\"{u}rzburg, Germany \\
$^{2}$A. F. Ioffe Physico-Technical Institute, Russian Academy of
Sciences, 194021 St. Petersburg, Russia}

\date{\today}

\begin{abstract}
We report a surprisingly long spin relaxation time of electrons in
Mn-doped p-GaAs. The spin relaxation time scales with the optical
pumping and increases from 12~ns in the dark to 160~ns upon
saturation. This behavior is associated with the difference in
spin relaxation rates of electrons precessing in the fluctuating
fields of ionized or neutral Mn acceptors, respectively. For the
latter the antiferromagnetic exchange interaction between a Mn ion
and a bound hole results in a partial compensation of these
fluctuating fields, leading to the enhanced spin memory.

\end{abstract}


\maketitle
%---------------------------------------------------------------

It has been known since the seventies \cite{Early}, that electrons
in n-doped GaAs possess a long spin lifetime. Because of high
electron mobilty, it is proved possible to transfer electron spin
over macroscopic distances (yielding a spin diffusion length of ca
10~$\mathrm{\mu m}$ \cite{Dzh1} and a spin drift length up to
100~$\mathrm{\mu m}$ \cite{SpinDrift1,SpinDrift2,SpinDrift3}).
Recent studies revealed a nontrivial doping concentration
dependence of the spin relaxation time $\tau_s$ \cite{Hanle1},
with  $\tau_s$ ranging up to 300~ns
\cite{Long-GaAs1,Long-GaAs2,Long-GaAs3}. In contrast, p-GaAs
exhibits a short electron spin relaxation time ($\tau_s \sim
1$~ns) and because of this cannot be regarded as technologically
promising. The fast electron spin relaxation in p-GaAs results
from the exchange interaction with holes bound to acceptors, which
is referred to as the Bir-Aronov-Pikus (BAP) mechanism
\cite{ref_SpinHoles, OO}. One might anticipate that doping of GaAs
with magnetic acceptors would result in even shorter spin
relaxation times than ordinarily observed in p-GaAs, because of
the additional spin scattering one expects from the Mn spins.

\begin{figure}[b]
\includegraphics[width=.39\textwidth]{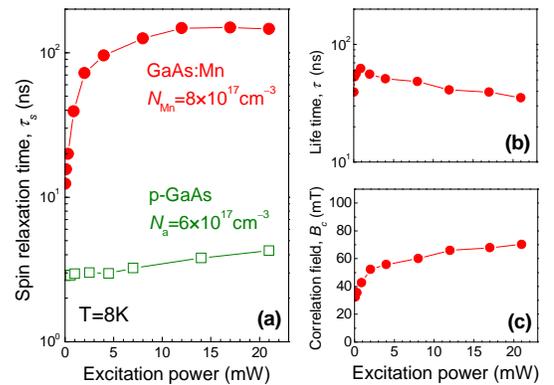}
\caption{(Color online) (a) Spin relaxation time $\tau_s$ vs
excitation power $P$ for GaAs:Mn ($N_{\mathrm{Mn}} = 8 \times 10
^{17} \, \mathrm{cm^{-3}}$) (solid circles) and for p-GaAs
containing non-magnetic acceptors ($N_{a} = 6 \times 10 ^{17} \,
\mathrm{cm^{-3}}$) (open squares). (b) Lifetime $\tau$  and (c)
correlation field $B_c$ vs $P$ for the GaAs:Mn sample.}
\label{fig1}
\end{figure}

Here we report on an intriguing and rather unexpected behavior of
the electron spin relaxation in Mn-doped GaAs. We find that doping
GaAs with Mn acceptors with concentration $N_{\mathrm{Mn}} \approx
10^{17} \, \mathrm{cm^{-3}}$ causes an increase of the electron
spin relaxation time by two orders of magnitude as compared with
similarly doped p-GaAs containing non-magnetic acceptors.
Figure~\ref{fig1}(a) shows that the electron spin relaxation time
in GaAs:Mn reaches 160~ns at elevated excitation power, which is
comparable with the best results achieved in n-GaAs
\cite{Long-GaAs1,Long-GaAs2,Long-GaAs3}.

The effect originates from a cancelation of the effective magnetic
fields acting on an electron by the antiferromagnetically aligned
hole and Mn spins. This cancelation drastically suppresses the
electron spin flip rate by the acceptors. Mathematically, the
contribution to $\tau_s$ arising from spin scattering can be
written as \cite{wf2_tc,OO}
\begin{equation}
\tau_s^{-1} = \frac{2}{3} \langle \omega_f^2 \rangle \tau_c \,.
\label{Eq1}
\end{equation}
Here, $\omega_f$ is the precession frequency of the electron spin
in the magnetic fields produced by a (magnetic) impurity and/or a
hole and $\tau_c$ is a correlation time that indicates how long
these fields remain unchanged, both in magnitude and direction.
When the electrons are localized on shallow donors, $\tau_c$ can
be interpreted as a characteristic electron hopping time.
Equation~(\ref{Eq1}) has a clear physical meaning. The higher the
precession frequency $\omega_f$, the faster the spin becomes
randomized, and the higher is the spin relaxation rate. Similarly,
for a short correlation time $\tau_c$, the fluctuating fields are
averaged out, and the spin relaxation time is longer. In GaAs:Mn,
due to the antiferromagnetic interaction between magnetic
impurities and holes \cite{Mn_1,ref_Mn-GaAs} their fluctuating
fields tend to compensate each another ($\omega_f \rightarrow 0$)
leading to a spin memory enhancement.

We have found a suppression of spin relaxation in various GaAs:Mn
structures with $N_{\mathrm{Mn}}$ ranging from $ 10 ^{17}$ to $10
^{18}$~$\mathrm{cm^{-3}}$. For clarity, we focus here on data from
one representative sample, grown by liquid-phase epitaxy (LPE) on
a (001)-oriented GaAs substrate. This sample has a 36-$\mathrm{\mu
m}$-thick GaAs:Mn layer with $N_{\mathrm{Mn}} = 8 \times 10 ^{17}
\, \mathrm{cm^{-3}}$. The p-doping is partially compensated due to
the presence of residual donors. A second sample serves as a
reference and has a 3.2-$\mathrm{\mu m}$-thick LPE-grown p-GaAs
layer doped with nonmagnetic acceptors (Ge) at a level $N_{a} = 6
\times 10 ^{17} \, \mathrm{cm^{-3}}$.

The electron spin dynamics have been studied by means of optical
orientation under continuous wave (cw) excitation. A net spin
polarization is created by circularly polarized excitation,
provided by a Ti-sapphire laser operating at 800~nm and focused to
a $300$-$\mathrm{\mu m}$ diameter spot on the sample. The
photoluminescence (PL) spectra are dispersed by a 1-m spectrometer
and detected by a Si-based avalanche photodiode. In order to
eliminate dynamic nuclear polarization, the helicity of the
excitation polarization is modulated at 50~kHz, and the
polarization of the emission $\rho_c = (I_{+}^{+} -
I_{+}^{-})/(I_{+}^{+} + I_{+}^{-})$ is detected using a
two-channel photon counter. Here, $I_{+}^{+}$ ($I_{+}^{-}$) refers
to the intensity of $\sigma^{+}$-polarized PL under $\sigma^{+}$
($\sigma^{-}$)-polarized excitation. Magnetic fields are applied
either perpendicular to, or in the sample plane (Faraday and Voigt
geometry, respectively). Most of the experiments have been
performed for a sample temperature of $T = 8$~K, unless otherwise
indicated.

\begin{figure}[tbp]
\includegraphics[width=.39\textwidth]{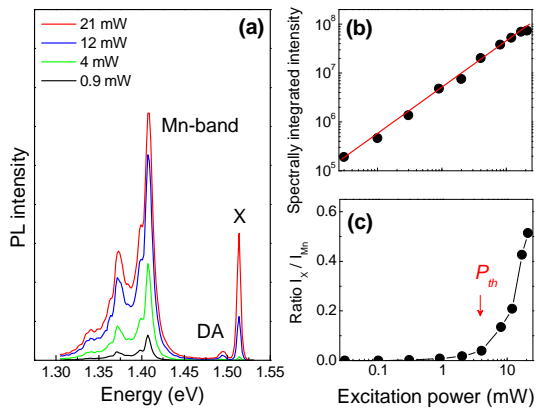}
\caption{(Color online) (a) PL spectra of the GaAs:Mn sample for
different excitation powers as indicated in the panel. (b) Spectrally
integrated PL intensity vs excitation power. (c) Intensity ratio
$I_{X} / I_{\mathrm{Mn}}$ of the excitonic (X) and Mn lines vs
excitation power.} \label{fig2}
\end{figure}

The PL spectra of the GaAs:Mn sample are shown in
Fig.~\ref{fig2}(a). The two lines at higher energies (1.51~eV and
1.49~eV) originate from excitonic (X) and shallow donor-acceptor
(DA) transitions and are similar to what is observed in
compensated, nonmagnetically-doped GaAs. The lower energy part of
the PL, peaking at 1.41~eV, is a Mn-related band,
spectrally-broadened by electron-phonon coupling. At low
temperatures this band can be attributed to transitions from
shallow donors to the Mn acceptor level, and their phonon replicas
\cite{Mn_PL}. Remarkably, the Mn-band is observed at all
excitation powers $P$, while the X and DA lines appear in the PL
spectra only when $P$ exceeds a threshold value $P_{th} \approx
5$~mW. This is evident also from Fig.~\ref{fig2}(c) where the
ratio $I_X / I_{\mathrm{Mn}}$ of the PL intensities detected at
1.51~eV ($I_{X}$) and 1.41~eV ($I_{\mathrm{Mn}}$) is plotted vs
$P$. For the same power interval, the spectrally integrated PL
intensity over the whole 1.3-1.55~eV energy range exhibits a
perfectly linear dependence on $P$ [Fig.~\ref{fig2}(b)]. This
indicates that the total PL intensity redistributes between the Mn
and X (DA) lines.

These data can be interpreted as follows: In partially compensated
GaAs:Mn, a manganese impurity occurs in the GaAs host either as a
neutral $A^0$ ($d^5 + h$) or ionized $A^-$ ($d^5$) acceptor center
\cite{ref_Mn-GaAs,Mn_1, Mn_PL, Mn_2,Mn_3}. Under excitation, $A^-$
centers trap photo-generated holes. Therefore, at low excitation
powers ($P < P_{th}$) the only radiative channel is the
recombination of electrons with holes bound to Mn ions. The
lifetime $\tau$ of an electron-hole pair is rather long (we find
$\tau \geq 40$~ns as discussed below) because electrons and holes
are localized on spatially separated centers. With growing $P$
this radiative channel saturates rapidly, such that for $P =
P_{th}$ all Mn impurities are converted into the $A^0$
configuration. With $P> P_{th}$ increases even further, a new
radiative channel opens up, due to the availability of
photo-excited valence-band holes. As a result, the X and DA lines
now start to appear in the PL spectra, still in addition to the
Mn-band.

We use the Hanle effect to measure the spin relaxation time
$\tau_s$ and the lifetime $\tau$ of optically excited electrons
\cite{OO}. Because of fast hole spin relaxation, the PL
polarization $\rho_{c}$ reflects the net spin polarization of the
electrons. In an external magnetic field $B$ applied in Voigt
geometry the electron spins precess with the Larmor frequency
$\Omega = g_{e} \mu_{B}B$ ($g_{e}$ is the electron g-factor)
around the field direction. As a result, the average spin (and
thus $\rho_c$) decreases with increasing $B$ as
\begin{equation}
\rho_c = \frac{\rho_{c0}}{1+ \tau / \tau_s} \frac{1}{1 +(B /
B_{1/2})^2} \,. \label{Eq2}
\end{equation}
Here, the characteristic field $B_{1/2} = \hbar / (g_e \mu_B)
T_s^{-1}$ gives the inverse spin lifetime $T_s^{-1} = \tau_s^{-1}
+ \tau^{-1}$. We take the g-factor of our samples to be equal to
that of pure GaAs (i.e., $g_e = -0.44$), as for the given
$N_{\mathrm{Mn}} = 8 \times 10 ^{17} \, \mathrm{cm^{-3}}$
(corresponds to $x=0.004 \%$) the correction to the g-factor due
to the $s$-$d$ exchange interaction ($| N_0 \alpha | \approx
0.17$~eV) is below 30\% \cite{g_GaAs1,g_GaAs2}. Thus, if the
'initial polarization' $\rho_{c0}$ is known (this can be estimated
from an experiment in Faraday geometry, see below), $\tau_s$ and
$\tau$ can be unambiguously determined from the Hanle curves.

\begin{figure}[tbp]
\includegraphics[width=.39\textwidth]{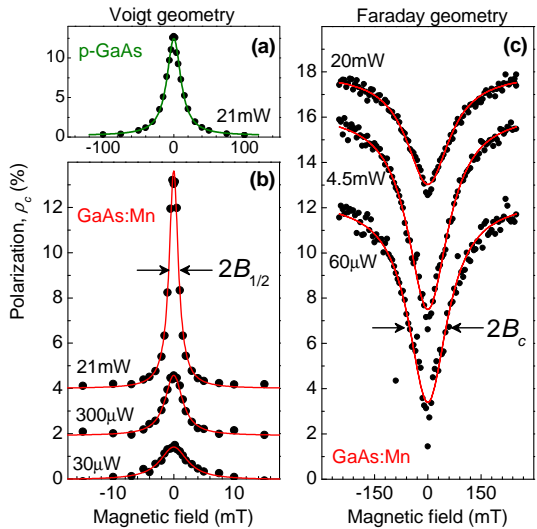}
\caption{(Color online) (a) A Hanle curve obtained from the
not-magnetically doped p-GaAs reference sample. (b) Hanle curves
from the GaAs:Mn sample for various excitation power levels
(indicated in the panel). (c) Recovery of the spin polarization in
an external magnetic field in Faraday geometry for $P$= 0.06, 4.5
and 20 mW. Solid lines in (a) and (b) are fits to Eq.~(\ref{Eq2}),
those in (c) are fits to Eq.~(\ref{Eq3}). In (b) and (c) the upper
curves are off-set vertically for clarity.} \label{fig3}
\end{figure}

Figure~\ref{fig3}(b) illustrates the dramatic changes of the Hanle
curves (detected at 1.41~eV ) in GaAs:Mn with increasing
excitation power $P$. From the fits to Eq.~(\ref{Eq2}) (indicated
by the solid lines in this Figure) we can determine the
dependencies $\tau_s (P)$ and $\tau (P)$. The lifetime $\tau$
[Fig.~\ref{fig1}(b)] is rather long and depends only slightly on
$P$, varying from 40~ns for small $P$ to 70~ns for $P=20$~mW. In
contrast, the spin relaxation time $\tau_s$ shows a very strong
dependence on excitation power [circles in Fig.~\ref{fig1}(a)]. It
increases from 12~ns for the lowest power $P= 30$~$\mathrm{\mu W}$
and saturates at about 160~ns for high power levels. Remarkably,
the saturation of $\tau_s$ occurs at a threshold value $P_{th}
\approx 5$~mW, very similar to the value of $P_{th}$ found from
the appearance of X and DA lines in the PL spectra (see
Fig.~\ref{fig2}). This implies that the spin relaxation time is
directly linked to the charge state of Mn acceptors and is long
when all Mn impurities are in $A^0$ configuration.

That the behavior of $\tau_s$ in GaAs:Mn is unusual is also
confirmed by a comparison with the reference sample.
Figure~\ref{fig3}(a) shows a Hanle curve detected at the DA line
in nonmagnetically doped p-GaAs [note that the field scale is
expanded by 10 compared to that in Fig.~\ref{fig3}(b)]. From fits
to Eq.~(\ref{Eq2}) [e.g., the solid line in Fig.~\ref{fig3}(a)] we
can determine $\tau_s$ in p-GaAs, assuming $\rho_{c0}=0.25$
\cite{OO}. The spin relaxation time in the reference sample turns
out to be much shorter [squares in Fig.~\ref{fig1}(a)] than in
GaAs:Mn.
% and nearly independent of $P$, increasing from 3~ns to 4~ns

We have performed additional experiments in order to analyze how
the correlation time $\tau_c$ [entering Eq.~(\ref{Eq1})] depends
on $P$. When an external magnetic field is applied in the Faraday
configuration, the transverse components of the fluctuating
internal fields that cause spin relaxation can be dynamically
averaged out. This occurs when $\Delta E_z \tau_c / \hbar \sim 1$
\cite{OO,Hanle1}, where, $\Delta E_z  = \mu_B (g_A - g_e) B$ and
$g_A$ depends on the electronic configuration of the Mn
accceptor($g_{A^0} = +2.77$ and $g_{A^-} = +2.00$
\cite{ref_Mn-GaAs}). As a result, the spin polarization recovers,
according to
\begin{equation}
\rho_c = \frac{\rho_{c0}}{1+ \tau / \tau^*_s}  \,,\,\,\,\,\,
\tau^*_s = \tau_s [1+(B / B_c)^2] \,, \label{Eq3}
\end{equation}
where $B_c = \hbar / [(g_A - g_e) \mu_B] \tau_c^{-1}$.
Figure~\ref{fig3}(c) shows the experimental data from the GaAs:Mn
sample (circles) together with fits (lines) to Eq.~(\ref{Eq3}). At
high fields $\rho_c$ tends to the initial polarization $\rho_c
\rightarrow \rho_{c0}$, and these values of $\rho_{c0}$ have been
used in fitting the Hanle curves discussed above. The correlation
field $B_c$ extracted from the dependencies in Fig.~\ref{fig3}(c)
are plotted as a function of the excitation power $P$ in
Fig.~\ref{fig1}(c). It only weakly depends on $P$, increasing from
36~mT to 70~mT. Assuming that in the low power limit most Mn
acceptors are in the $A^-$ configuration and that above $P_{th}$
all Mn are in the $A^0$ configuration, we find correlation times
$\tau_c^{A^-} = 131$~ps and $\tau_c^{A^0} = 52$~ps, respectively.
While qualitatively these correlation times show the correct
behavior to explain the observed increase of $\tau_s$ with $P$,
their ratio $\tau_c^{A^-} / \tau_c^{A^0} = 2.5$, when plugged into
Eq.~(\ref{Eq1}) is insufficient to explain the 13-fold increase of
$\tau_s$ in Fig.~\ref{fig1}(a).

Hence, Eq.~(\ref{Eq1}) suggests that the suppression of spin
relaxation in GaAs:Mn is caused by a reduction of the amplitude of
the fluctuating fields when the Mn acceptors are in the $A^0$
configuration. This arises from the antiferromagnetic exchange
interaction ($\Delta^{A^{0}}_{pd} < 0$) between the Mn d-shell
electrons (with spin $S_d = 5/2$) and the hole (with quasi-spin
$J_h = 3/2$)
\begin{equation}
\mathcal{\widehat{H}}^{A^{0}}_{pd} = - \Delta^{A^0}_{pd} \,
\mathbf{\widehat{S}_d} \mathbf{\widehat{J}_h} \,, \label{Eq4}
\end{equation}
where the exchange integral $\Delta^{A^0}_{pd} = -2.2$~meV
\cite{g_GaAs1}. $\mathcal{\widehat{H}}^{A^{0}}_{pd}$ splits the
$A^0$ state into four sublevels characterized by angular momentum
$F = | S_d + J_h | = 1,2,3,4$. In its ground state ($F = 1$), the
Mn and hole spins are oriented antiparallel, and the fluctuating
fields associated with them compensate one another.

We now estimate the influence this effect has on $\tau_s$. First, consider the
exchange interaction between an electron with spin $S_e = 1/2$
bound to a shallow donor (we assume a Bohr radius $a_B = 100$~{\AA}) and
an ionized $A^-$ center at a distance $R_d$
\begin{equation}
\mathcal{\widehat{H}}^{e}_{sd} = -b \, \mathbf{\widehat{S}_d}
\mathbf{\widehat{S}_e} \,, \label{Eq5}
\end{equation}
where $b = b_0 \exp(-2 R_d / a_B)$. The parameter $b_0$ can be
estimated from $\alpha N_0 \approx 0.17$~eV \cite{g_GaAs1}
normalized by the number of cations $4 / a_0^3$ ($a_0 = 5.65$~{\AA}
is the lattice constant) in a localization volume $\pi a_B^3$.
This gives $b_0 = \alpha N_0  a_0^3 / (4 \pi a_B^3) \approx
2.3$~$\mathrm{\mu eV}$. Thus the precession
frequency of the electron spin in the fluctuating field of the ionized Mn acceptor can
be written as

\begin{equation}
\omega_f^{A^-} = b \sqrt{S_d (S_d +1)} / \hbar = b
\frac{\sqrt{35}}{2 \hbar} \,. \label{Eq6}
\end{equation}

When the remote acceptor is in the $A^0$ state, also the electron-hole exchange interaction has
to be considered:
\begin{equation}
\mathcal{\widehat{H}}^{e}_{sp} = -\delta \, \mathbf{\widehat{J}_h}
\mathbf{\widehat{S}_e} \,. \label{Eq7}
\end{equation}
Here, $\delta = \delta_0 \exp(-2 R_d / a_B)$ and the parameter
$\delta_0$ can be estimated from the exciton exchange splitting in
bulk GaAs ($2 \delta_0 \approx 6$~$\mathrm{\mu eV}$ \cite{eh_GaAs}). The
total three particle exchange Hamiltonian thus takes the form
\begin{equation}
\mathcal{\widehat{H}}^{e-A^0}_{exch} = \mathcal{\widehat{H}}^{A^{0}}_{pd} +
\mathcal{\widehat{H}}^{e}_{sd} + \mathcal{\widehat{H}}^{e}_{sp} \,.
\label{EqSum}
\end{equation}
Because $\Delta^{A^{0}}_{pd} \gg b, \delta$ the last two terms can be
considered as a perturbation. In other words, only the components
% $\mathbf{F} (\mathbf{S_d} \mathbf{F} )/ (F^2)$ and $\mathbf{F}
% (\mathbf{J_h} \mathbf{F} )/ (F^2)$
of the vectors $\mathbf{S_d}$ and $\mathbf{J_h}$ along the
direction of $\mathbf{F} = \mathbf{S_d} + \mathbf{J_h}$ are
preserved. Within a subspace of given $F^2$, $S_d^2$ and $J_h^2$
the spin Hamiltonian now becomes
\begin{equation}
\mathcal{\widehat{H}}^{e-A^0}_{exch} = - \frac{\Delta^{A^{0}}_{pd}}{2} ( \widehat{F}^2 -
\widehat{S}_d^2 - \widehat{J}_h^2) - a_F \mathbf{\widehat{F}}
\mathbf{\widehat{S}_e} \,. \label{Eq8}
\end{equation}
Here, $a_F$
% $a_F = b \langle \mathbf{\widehat{S}_d} \mathbf{\widehat{F}}
% \rangle / \langle \mathbf{\widehat{F}}^2 \rangle + \delta \langle
% \mathbf{\widehat{J}_h} \mathbf{\widehat{F}} \rangle / \langle
% \mathbf{\widehat{F}}^2 \rangle$
can be found in the usual manner for coupling angular momentum
\cite{Sum_M}, yielding
\begin{equation}
a_F = \frac{1}{2} [(\delta + b) + (\delta -b) \frac{J_h (J_h +1) -
S_d (S_d+1)}{F (F+1)}] \,. \label{Eq9}
\end{equation}
Finally, by analogy with Eq.~(\ref{Eq5}) we obtain the precession
frequency of the electron in the fluctuating field of the neutral
Mn acceptor [which we assume to be in the ground state ($F = 1$)]:
\begin{equation}
\omega_f^{A^0} = | a_F | \sqrt{F (F +1)} / \hbar =  \frac{| 7 b -
3 \delta |}{2 \sqrt{2} \hbar}\,. \label{Eq10}
\end{equation}
Obviously, for $\delta / b= 7 / 3$ the fluctuating field is
screened out completely. From the data for $\tau_s$ and
$\tau_c$ presented in Fig.~\ref{fig1} and using Eq.~(\ref{Eq1})
one obtains $ ( \omega_f^{A^-} / \omega_f^{A^0} )^2
\approx 5$. This ratio, using Eqs.~(\ref{Eq6}) and (\ref{Eq10}),
implies we have $\delta / b = 2.3$ in our sample. This agrees remarkably well with
our rough estimates given above, which give $\delta / b = 1.3$.

\begin{figure}[tbp]
\includegraphics[width=.37\textwidth]{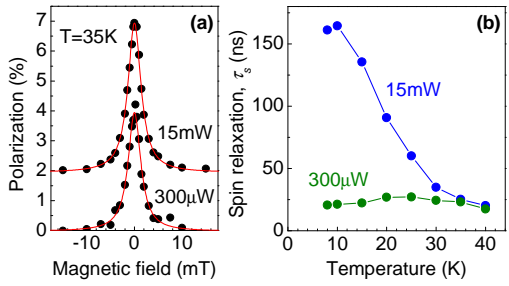}
\caption{(Color online) (a) Hanle curves for two excitation powers
(shown in the panel) measured at a temperature $T = 35$~K. (b)
Spin relaxation times $\tau_s$ vs $T$ for the same excitation
power levels as in (a). Lines in (b) are guides to an eye.}
\label{fig5}
\end{figure}

Further evidence for the correctness of our model comes from
temperature dependent studies.  With increasing temperature the
antiferromagnetic exchange correlation between Mn ions and holes
is washed out. Thus their respective fluctuating fields cannot
compensate each another any more, and as a result one expects that
$\tau_s$ should decrease strongly when the sample mainly contains
$A^0$ centers (i.e., in the high excitation regime). At the same
time, an increase in temperature should have little effect on
$\tau_s$ when the electron spins are predominantly scattered by
$A^-$ centers (at low excitation levels). This behavior is indeed
observed experimentally (Fig.~\ref{fig5}). At $T = 35$~K, the
Hanle curves for low ($P = 300$~$\mathrm{\mu W}$) and high ($P =
15$~mW) excitation powers [Fig.~\ref{fig5}(a)] are very similar,
in strong contrast with the behavior at 8~K [Fig.~\ref{fig3}(b)].
The full temperature dependence of $\tau_s$ for these excitation
levels is shown in Fig.~\ref{fig5}(b). In agreement with our
expectations, $\tau_s$ is nearly constant for $P =
300$~$\mathrm{\mu W}$ (predominantly $A^-$ centers), while it is
strongly reduced for $P = 15$~mW , where the sample contains
mainly $A^0$ centers. We observed that the spin relaxation
activates with a characteristic energy of ca $k_B T_0 = 2.2$~meV,
where $T_0 = 25$~K, which correlates well with the exchange energy
$\Delta^{A^0}_{pd} = 2.2$~meV.

In closing, we note that the observed reduction of spin relaxation by
exchange coupling on a magnetic dopant may open up a new route
for spin memory engineering.

This research was supported by the DFG (436 RUS 113/843 and SPP
1285) as well as the RFBR.

%---------------------------------------------------------------
%***************************************************************
%---------------------------------------------------------------

\end{document}